\documentstyle[12pt]{article}
\newcommand{\text}{\rm}

\begin{document}

\title{{\bf Vector Supersymmetry of 2D Yang-Mills Theory}}
\author{{\bf J.L. Boldo$^{*}$, C.A.G. Sasaki$^{\dagger }$, S.P. Sorella$^{\dagger
\dagger }$, } \and {\bf and L.C.Q. Vilar$^{\dagger \dagger }$,}\vspace{2mm} 
\\
{\bf $^{\ast}$}CBPF, Centro Brasileiro{\bf \ }de Pesquisas F\'{\i}sicas \\
Rua Xavier Sigaud 150, 22290-180, Urca, \\
Rio de Janeiro, Brazil.\vspace{2mm}\\
{\bf $^{\dagger }$}UERJ, Universidade do Estado do Rio de Janeiro,\\
Departamento de Estruturas Matem\'{a}ticas, \\
Instituto de Matem\'{a}tica e Estat\'{\i}stica, \\
Rua S\~{a}o Francisco Xavier, 524, \\
20550-013, Maracan\~{a}, Rio de Janeiro, Brazil.\vspace{2mm}\\
{\bf $^{\dagger \dagger }$}UERJ, Universidade do Estado do Rio de Janeiro,\\
Departamento de F\'{\i}sica Te\'{o}rica, \\
Instituto de F\'{\i}sica, \\
Rua S\~{a}o Francisco Xavier, 524, \\
20550-013, Maracan\~{a}, Rio de Janeiro, Brazil.\vspace{2mm}}
\maketitle

\begin{abstract}
The vector supersymmetry of the 2D topological $BF$ model is extended to 2D
Yang-Mills. The consequences of the corresponding Ward identity on the
ultraviolet behavior of the theory are analyzed.

\setcounter{page}{0}\thispagestyle{empty}
\end{abstract}

\vfill\newpage\ \makeatother
\renewcommand{\theequation}{\thesection.\arabic{equation}} %
\renewcommand{\baselinestretch}{2}

\section{\ Introduction}

The relation between the two-dimensional Yang-Mills theory (2DYM) and the
topological models has been object of intensive investigations over the past
years \cite{Cordes,Blau,Billo,soda}. In spite of the lacking of local
degrees of freedom, 2DYM is nontrivial when analyzed from the point of view
of the topological field theories, as underlined for instance by \cite{soda}
within the BRST framework.

An interesting feature of the topological theories, of both Witten and
Schwartz type, is the existence, besides their BRST symmetry, of an
additional invariance whose generators carry a vector index. This further
symmetry, called vector supersymmetry \cite{BRT e BR,DGS e DLPS,GMS}, gives
rise, together with the BRST generator, to an algebra of the Wess-Zumino
type which, closing on-shell on the space-time translations, allows for a
supersymmetric interpretation of the topological models \cite{BRT e BR,DGS e
DLPS,GMS}. In particular, it has been shown \cite{algebraic} that the
existence of the vector supersymmetry is deeply related to the fact that the
energy-momentum tensor can be expressed in the form of a pure BRST
variation, a key feature which can be taken as the proper definition of the
topological theories.

It seems therefore natural to ask ourselves whether this supersymmetric
structure can also be found in the 2DYM; this being the goal of the present
letter. As one can easily guess, we will be able to show that this question
can be actually answered in the affirmative. As a by-product, a simple
understanding of the ultraviolet finiteness of 2DYM will be provided.

\section{Symmetries of 2DYM}

Let us start by considering the gauge invariant Yang-Mills action in $2D$
euclidean space-time:

\begin{equation}
S_{YM}=-\frac 14\int d^2x\,\,F^{a\mu \nu }F_{\mu \nu }^a\;,  \label{in-act}
\end{equation}
where $F_{\mu \nu }^a=\partial _\mu A_\nu -\partial _\nu A_\mu
+gf^{abc}A_\mu ^bA_\nu ^c$ is the field strength and $g$ stands for the
gauge coupling constant. Notice that in $2D$ the gauge connection is
dimensionless, so that the coupling constant $g$ has dimension one.

In order to analyze the symmetries of this model, it is convenient to switch
to the first order formalism \cite{soda} by rewriting the expression (\ref
{in-act}) in the following form 
\begin{equation}
S_{YM}=S_{top}+S_\phi \;,  \label{2dym}
\end{equation}
where

\begin{equation}
S_{top}=\frac 12\int d^2x\,\,\varepsilon ^{\mu \nu }F_{\mu \nu }^a\phi ^a\;,
\label{top}
\end{equation}
and 
\begin{equation}
S_\phi =\frac 12\int d^2x\,\,\phi ^a\phi ^a,  \label{phi}
\end{equation}
$\phi ^a$ being an auxiliary scalar field. Expression (\ref{2dym}) is
obviously seen to be equivalent to (\ref{in-act}) upon elimination of the
auxiliary field $\phi ^a$ through the equations of motion. It worth
remarking here that the use of the first order formalism allows us to
interpret to some extent the 2DYM as a deformation of a topological field
theory, as it is easily recognized that the term $S_{top}$ in eq.(\ref{top})
is the action of the two-dimensional topological $BF$ model \cite{bm}. The
second term $S_\phi $ is metric-dependent, playing therefore the role of the
deformation. Expression (\ref{2dym}) also suggests what will our strategy be
in order to establish the vector susy Ward identity for the 2DYM. To this
end we recall that the first term $S_{top}$ of eq.(\ref{2dym}), identifying
a topological field theory, possesses the vector supersymmetry \cite{bm}
which, however, will not leave the second term invariant. Nevertheless, it
is a remarkable property of the 2DYM that the breaking terms stemming from
the non-invariance of the metric-dependent part $S_\phi $ of the action (\ref
{2dym}) can be taken into account by the introduction of a suitable set of
external fields. As we shall see, this procedure will enable us to write
down an off-shell Ward identity which is an extension of the vector susy
Ward identity of the topological two-dimensional $BF$ model \cite{bm}. This
identity will strongly constrain the ultraviolet behavior of the 2DYM.

Once established the classical counterpart of the theory, our next step is
to get its quantum version. To this aim one has to fix the gauge invariance
of the action; we add the following gauge fixing term, by using the Landau
gauge condition: 
\begin{equation}
S_{gf}=\int d^2x\,\,\left( b^a\partial ^\mu A_\mu ^a-\partial ^\mu \overline{%
c}^a(D_\mu c)^a\right) {\rm {\ },}  \label{landau}
\end{equation}
where $c$, $\overline{c}^a$ and $b^a$ are the ghost, the anti-ghost and the
Lagrange multiplier, respectively. The gauge fixed action 
\begin{equation}
S=S_{YM}+S_{gf}\;=S_{top}+S_\phi \;+S_{gf}\;,  \label{g-f}
\end{equation}
is invariant under the BRST transformations: 
\begin{equation}
\begin{array}{l}
sA_\mu ^a=-\left( \partial _\mu c^a+gf^{abc}A_\mu ^bc^c\right) , \\ 
\\ 
sc^a=\frac 12gf^{abc}c^bc^c, \\ 
\\ 
s\phi ^a=gf^{abc}c^b\phi ^c, \\ 
\\ 
s\overline{c}^a=b^a,\,\,sb^a=0\;.
\end{array}
\label{BRS-transf}
\end{equation}
The dimension and the ghost-number of the fields are displayed in Table 1.

\begin{center}
\[
\begin{tabular}{|c|c|c|c|c|c|}
\hline
& $A_\mu ^a$ & $\phi ^a$ & $\,\,\,\,\,\,\,\,c^a$ & $\bar{c}^a$ & $b^a$ \\ 
\hline
$Dim$ & $0$ & $1$ & $0$ & $0$ & $1$ \\ \hline
$Ng$ & $0$ & $0$ & $1$ & $-1$ & $0$ \\ \hline
\end{tabular}
\]

Table 1
\end{center}

\noindent Let us now focus on the following sector of the action (\ref{g-f}%
): 
\begin{equation}
S_{inv}=S_{top}+S_{gf}\;,
\end{equation}
corresponding to the quantized topological $BF$ model$.$ As already
underlined, besides the BRST invariance (\ref{BRS-transf}), $S_{inv}$ has a
further symmetry; the so called vector supersymmetry \cite{bm}, which reads 
\begin{equation}
\begin{array}{l}
\delta _\mu A_\nu ^a=0\;, \\ 
\\ 
\delta _\mu \phi ^a=-\varepsilon _{\mu \nu }\partial ^\nu \overline{c}^a, \\ 
\\ 
\delta _\mu c^a=-A_\mu ^a\;, \\ 
\\ 
\delta _\mu \overline{c}^a=0\;, \\ 
\\ 
\delta _\mu b^a=\partial _\mu \overline{c}^a.
\end{array}
\label{susy}
\end{equation}
In summary, we have 
\begin{equation}
\begin{array}{l}
sS_{inv}=0\;, \\ 
\delta _\mu S_{inv}=0\;.
\end{array}
\label{bd-eq}
\end{equation}
In addition, the generators $s$ and $\delta _\mu $ give rise to the
following Wess-Zumino supersymmetric algebra

\begin{eqnarray}
\left\{ s,\delta _\mu \right\} \phi ^a &=&\partial _\mu \phi ^a+\varepsilon
_{\mu \nu }\frac{\delta S_{inv}}{\delta A_\nu ^a}\;,  \label{v-s} \\
&&  \nonumber \\
\left\{ s,\delta _\mu \right\} A_\nu ^a &=&\partial _\mu A_\nu
^a-\varepsilon _{\mu \nu }\frac{\delta S_{inv}}{\delta \phi ^a}\;,  \nonumber
\\
&&  \nonumber \\
\left\{ s,\delta _\mu \right\} \left( c,b,\overline{c}\right) &=&\partial
_\mu \left( c,b,\overline{c}\right) \;,  \nonumber
\end{eqnarray}
which, closing on-shell on the space-time translations, allows for a
supersymmetric interpretation of the two-dimensional $BF$ model.

On the other hand, we can easily verify that, as expected, the
metric-dependent term $S_\phi $ of the quantized 2DYM action in eq.(\ref{g-f}%
) breaks the vector supersymmetry invariance (\ref{susy}) of the topological
sector. In fact 
\begin{equation}
\delta _\mu {\cal O}=-\varepsilon _{\mu \nu }\phi ^a\partial ^\nu \overline{c%
}^a,  \label{br}
\end{equation}
where we have defined ${\cal O}=\frac 12\phi ^a\phi ^a.$ As previously
remarked, our procedure in order to control the effects of this breaking is
to introduce external fields as follows:

\begin{equation}
S_{{\cal O}}=\int d^2x\,\,\left( \tau {\cal O}+\xi ^\mu \delta _\mu {\cal O}%
+\frac 12\eta ^{\mu \nu }\delta _\mu \delta _\nu {\cal O}+\frac 12\Omega
^{\mu \nu }s\delta _\mu \delta _\nu {\cal O}\right) \;,  \label{sext}
\end{equation}
with $\tau $ and $\xi ^\mu $ being respectively scalar and vector external
fields, while $\eta ^{\mu \nu }$ and $\Omega ^{\mu \nu }$ are rank-2
anti-symmetric external fields. Their canonical dimensions and ghost-numbers
are displayed in Table 2.

\begin{center}
$
\begin{tabular}{|c|c|c|c|c|c|c|}
\hline
& $s$ & $\delta _\mu $ & $\tau $ & $\xi _\mu $ & $\eta _{\mu \nu }$ & $%
\Omega _{\mu \nu }$ \\ \hline
$N_g$ & $1$ & $-1$ & $0$ & $1$ & $2$ & $1$ \\ \hline
$Dim$ & $1$ & $0$ & $0$ & $0$ & $0$ & $-1$ \\ \hline
\end{tabular}
$

Table 2
\end{center}

\noindent It follows thus that the action 
\begin{equation}
S_\tau =S+S_{{\cal O}}\;,  \label{Stau}
\end{equation}
is invariant under the modified nilpotent BRST transformations: 
\begin{equation}
\begin{array}{l}
\tilde{s}A_\mu ^a=-\left( D_\mu c\right) ^a+\xi ^\nu \varepsilon _{\nu \mu
}\phi ^a, \\ 
\\ 
\tilde{s}c^a=\frac 12gf^{abc}c^bc^c, \\ 
\\ 
\tilde{s}\phi ^a=gf^{abc}c^b\phi ^c, \\ 
\\ 
\tilde{s}\bar{c}^a=b_{a\;},\,\,\;\;\;\;\;\;\;\;\tilde{s}b_a=0\;, \\ 
\\ 
\tilde{s}\tau =-\partial _\mu \xi ^\mu ,\,\,\;\;\;\;\;\;\tilde{s}\xi _\mu
=0\;, \\ 
\\ 
\tilde{s}\Omega _{\mu \nu }=-\eta _{\mu \nu \;},\,\,\;\;\;\;\tilde{s}\eta
_{\mu \nu }=0\;,
\end{array}
\label{m-brst}
\end{equation}

\begin{equation}
\tilde{s}S_\tau =0\;.  \label{nilp}
\end{equation}
Moreover, remarkably in half, $S_\tau $ turns out to be left invariant by
the following extended susy transformations

\begin{eqnarray}
\widetilde{\delta }_\mu c^a\; &=&-A_\mu ^a\;,\;\;\;\;\;\;\;\;\;\;\;%
\widetilde{\delta }_\mu A_\nu ^a=0\;,  \label{ext-sv} \\
\widetilde{\delta }_\mu \phi ^a\; &=&-\varepsilon _{\mu \nu }\partial ^\nu 
\overline{c}^a,\;\;\;\;\;\;\;\widetilde{\delta }_\mu \overline{c}^a=0\;, 
\nonumber \\
\widetilde{\delta }_\mu b^a\; &=&\partial _\mu \overline{c}^a\;,  \nonumber
\\
\widetilde{\delta }_\mu \xi _\nu \; &=&-\delta _{\mu \nu }\left( 1+\tau
\right) \;,\;\;\;\;\widetilde{\delta }_\mu \tau =0\;, \\
\;\widetilde{\delta }_\mu \eta _{\nu \kappa } &=&-\delta _{\mu \kappa }\xi
_\nu +\delta _{\mu \nu }\xi _\kappa -\partial _\mu \Omega _{\nu \kappa \;}, 
\nonumber \\
\widetilde{\delta }_\mu \Omega _{\nu \kappa } &=&0\;,  \nonumber
\end{eqnarray}

\begin{equation}
\widetilde{\delta }_\mu S_\tau =0\;,  \label{ext-v-inv}
\end{equation}
where $\delta _{\mu \nu }$ is the flat euclidean metric. We see therefore
that, as announced, we have been able to account for the breaking generated
by the non-topological action $S_\phi $ by introducing suitable external
fields. We are now ready to implement the $\tilde{s}$ and the $\widetilde{%
\delta }_\mu -$invariance of the action $S_\tau $ as Ward identities. This
will be the task of the next section.

\section{Ward Identities}

Following the standard BRST procedure \cite{ps}, we add to expression (\ref
{Stau}) a new term $S_{ext}$, accounting for the nonlinear part of the
modified BRST transformations (\ref{m-brst}): 
\begin{equation}
S_{ext}=\int d^2x\,\,\left( \gamma ^{a\mu }\tilde{s}A_\mu ^a+L^a\tilde{s}%
c^a+\rho ^a\tilde{s}\phi ^a\right) \;,  \label{nl}
\end{equation}
where the external fields $\gamma ^{a\mu }$, $L^a$ and $\rho ^a$ have
dimensions and ghost-numbers displayed in Table 3.

\begin{center}
$%
\begin{tabular}{|l|l|l|l|}
\hline
& $\gamma ^{a\mu }$ & $L^a$ & $\rho ^a$ \\ \hline
$Dim$ & $1$ & $1$ & $0$ \\ \hline
$Ng$ & $-1$ & $-2$ & $-1$ \\ \hline
\end{tabular}
$

Table 3
\end{center}

\noindent Therefore, the complete action

\begin{equation}
\Sigma =S_\tau +S_{ext\;},  \label{c-a}
\end{equation}
obeys the Slavnov-Taylor identity

\begin{eqnarray}
{\cal S(}\Sigma ) &=&\int d^2x\,\,\left( \frac{\delta \Sigma }{\delta A_\mu
^a}\frac{\delta \Sigma }{\delta \gamma ^{a\mu }}+\frac{\delta \Sigma }{%
\delta \phi ^a}\frac{\delta \Sigma }{\delta \rho ^a}+\frac{\delta \Sigma }{%
\delta L^a}\frac{\delta \Sigma }{\delta c^a}\right.  \label{s-t} \\
\;\;\; &&\;\;\;\;\;\;\left. +b_a\frac{\delta \Sigma }{\delta \bar{c}^a}%
-\partial _\mu \xi ^\mu \frac{\delta \Sigma }{\delta \tau }-\frac 12\eta
^{\mu \nu }\frac{\delta \Sigma }{\delta \Omega ^{\mu \nu }}\right) =0\;. 
\nonumber
\end{eqnarray}
Let us also introduce, for further use, the nilpotent linearized
Slavnov-Taylor operator ${\cal B}_\Sigma $ 
\begin{eqnarray}
{\cal B}_\Sigma &=&\int d^2x\,\,\left( \frac{\delta \Sigma }{\delta A_\mu ^a}%
\frac \delta {\delta \gamma ^{a\mu }}+\frac{\delta \Sigma }{\delta \gamma
^{a\mu }}\frac \delta {\delta A_\mu ^a}+\frac{\delta \Sigma }{\delta \phi ^a}%
\frac \delta {\delta \rho ^a}+\frac{\delta \Sigma }{\delta \rho ^a}\frac
\delta {\delta \phi ^a}\right.  \label{lin-b} \\
&&+\left. \frac{\delta \Sigma }{\delta c^a}\frac \delta {\delta L^a}+\frac{%
\delta \Sigma }{\delta L^a}\frac \delta {\delta c^a}+b_a\frac \delta {\delta 
\bar{c}^a}-\partial _\mu \xi ^\mu \frac \delta {\delta \tau }-\frac 12\eta
^{\mu \nu }\frac \delta {\delta \Omega ^{\mu \nu }}\right) \;,  \nonumber
\end{eqnarray}

\begin{equation}
{\cal B}_\Sigma {\cal B}_\Sigma =0\;.  \label{nil-b}
\end{equation}
Turning now to the vector invariance $\widetilde{\delta }_\mu $, it is
easily verified that, due to the introduction of the external fields $\gamma
^{a\mu }$, $L^a$, $\rho ^a$, it takes the form of a linearly broken Ward
identity, namely 
\begin{equation}
{\cal W}_\mu \Sigma =\Delta _\mu ^{cl}\;,  \label{w-i}
\end{equation}
with 
\begin{eqnarray}
{\cal W}_\mu &=&\int d^2x\,\,\left( \varepsilon _{\mu \nu }\rho ^a\frac
\delta {\delta A_\nu ^a}-A_\mu ^a\frac \delta {\delta c^a}-\varepsilon _{\mu
\nu }\left( \gamma ^{a\nu }+\partial ^\nu \bar{c}^a\right) \frac \delta
{\delta \phi ^a}-L^a\frac \delta {\delta \gamma ^{a\mu }}\right.  \nonumber
\label{w-o} \\
&&+\left. \partial _\mu \bar{c}^a\frac \delta {\delta b^a}-\left( 1+\tau
\right) \frac \delta {\delta \xi ^\mu }+\frac 12\left( \delta _\mu ^\alpha
\xi ^\beta -\delta _\mu ^\beta \xi ^\alpha -\partial _\mu \Omega \right)
\frac \delta {\delta \eta ^{\alpha \beta }}\right) \;,\;  \nonumber
\label{w-o} \\
&&  \label{w-o}
\end{eqnarray}
and 
\begin{eqnarray}
\Delta _\mu ^{cl} &=&\int d^2x\,\left( -\gamma ^{a\nu }\partial _\mu A_\nu
^a+L^a\partial _\mu c^a-\rho ^a\partial _\mu \phi ^a\right.  \label{c-b} \\
&&-\left. \varepsilon _{\mu \nu }\rho ^a\partial ^\nu b^a-L^a\xi ^\nu
\varepsilon _{\nu \mu }\phi ^a+\gamma _\mu ^a\xi _\nu \partial ^v\bar{c}%
^a\right.  \nonumber \\
&&-\left. \gamma _\nu ^a\xi ^\nu \partial _\mu \bar{c}^a+\gamma _\nu
^a\gamma _\mu ^a\xi ^\nu \right) \;,  \nonumber
\end{eqnarray}
We observe that the breaking term $\Delta _\mu ^{cl}$, being linear in the
quantum fields, is a purely classical breaking and will not get renormalized 
\cite{ps}. It is worth underlining that the final form of the Ward identity (%
\ref{w-i}) is the expected one, being indeed a common feature of the
topological models, including in particular the associated classical
breaking term \cite{DGS e DLPS,GMS, algebraic,bm,wi}. We also point out
that, in the present case, the complete action $\Sigma $ is constrained by a
further linearly broken local Ward identity:

\begin{equation}
{\cal F}_{\mu \nu }(x)\Sigma =\Delta _{\mu \nu }^{{\cal F}}(x)\;,
\label{f-i}
\end{equation}
with

\begin{eqnarray}
{\cal F}_{\mu \nu }(x) &=&\varepsilon _{\nu \mu }\phi ^a\left( x\right)
\frac \delta {\delta c^a\left( x\right) }+\varepsilon _{\nu \mu }L^a\left(
x\right) \frac \delta {\delta \rho ^a\left( x\right) }  \label{f-o} \\
&&+\frac \delta {\delta \Omega ^{\mu \nu }\left( x\right) }+\partial _\nu
\frac \delta {\delta \xi ^\mu \left( x\right) }-\partial _\mu \frac \delta
{\delta \xi ^\nu \left( x\right) }  \nonumber \\
&&+\left( \gamma ^{av}+\partial ^v\bar{c}^a\right) \frac \delta {\delta
A^{a\mu }\left( x\right) }-\left( \gamma ^{a\mu }+\partial ^\mu \bar{c}%
^a\right) \frac \delta {\delta A^{a\nu }}\;,  \nonumber
\end{eqnarray}
and

\begin{equation}
\Delta _{\mu \nu }^{{\cal F}}(x)=\gamma _\mu ^a\partial _\nu b^a-\gamma _\nu
^a\partial _\mu b^a\;.  \label{f-b}
\end{equation}
In particular, the operators ${\cal B}_\Sigma ,$ ${\cal W}_\mu $ and ${\cal F%
}_{\mu \nu }(x)$ give rise to a closed algebra given by

\begin{eqnarray}
\left\{ {\cal B}_\Sigma ,{\cal W}_\mu \right\} &=&{\cal P}_\mu +\int d^2x\xi
^\nu {\cal F}_{\nu \mu }(x)\;,  \label{c-a} \\
\left\{ {\cal W}_\mu ,{\cal W}_\nu \right\} &=&\left\{ {\cal W}_\mu ,{\cal F}%
_{\sigma \rho }(x)\right\} =0\;,  \nonumber \\
\left\{ {\cal B}_\Sigma ,{\cal F}_{\sigma \rho }(x)\right\} &=&\left\{ {\cal %
F}_{\mu \nu }(x),{\cal F}_{\sigma \rho }(y)\right\} =0\;,  \nonumber
\end{eqnarray}
where ${\cal P}_\mu $ is the functional operator of the space-time
translations, {\it i.e.}

\begin{equation}
{\cal P}_\mu =\sum_{{\rm {(All\,fields\,}\varphi {)}}}\,\,\int
d^2x\,\,\partial _\mu \varphi \frac \delta {\delta \varphi }  \label{trans}
\end{equation}
Finally, let us display the whole set of conditions which are usually
imposed in the quantization of Yang-Mills theories in the Landau gauge \cite
{ps}, namely:

\vspace{5mm}

\noindent $\bullet \;$the linearly broken ghost equation Ward identity 
\begin{equation}
{\cal G}^a\Sigma =\Delta ^a\;,  \label{g-i}
\end{equation}
with 
\begin{equation}
{\cal G}^a=\int d^2x\,\left( \frac \delta {\delta c^a}+gf^{abc}\bar{c}%
^b\frac \delta {\delta b^c}\right) \;,  \label{g-o}
\end{equation}
and 
\begin{equation}
\Delta ^a=\int d^2x\,g\,f^{abc}\left( \gamma ^{b\mu }A_\mu ^c-L^bc^c+\rho
^b\phi ^c\right) \;.  \label{g-b}
\end{equation}
$\bullet \;$The gauge fixing condition 
\begin{equation}
\frac{\delta \Sigma }{\delta b^a}=\partial _\mu A^{a\mu }+\partial _\nu
\Omega ^{\mu \nu }\partial _\mu \bar{c}^a\;.  \label{l-c}
\end{equation}
$\bullet \;$The antighost equation 
\[
\left( \frac \delta {\delta \bar{c}^a}+\partial _\mu \frac \delta {\delta
\gamma _\mu ^a}\right) \Sigma =\partial _\nu \eta ^{\mu \nu }\partial _\mu 
\bar{c}^a+\partial _\nu \Omega ^{\mu \nu }\partial _\mu b^a. 
\]
As we shall see in the next section, the identities (\ref{s-t}), (\ref{w-i})
and (\ref{f-i}) turn out to have far-reaching consequences, accounting for
instance for the absence of nontrivial invariant counterterms.

\section{Invariant Counterterms}

Following the set up of the algebraic renormalization \cite{ps} and making
use of the general results on the cohomology of the Yang-Mills theories \cite
{mh}, it is not difficult to establish that the model and its Ward
identities are renormalizable. Here, we shall limit ourselves only to state
the final result, aiming to provide an algebraic understanding of the
finiteness properties of 2DYM. Let us look then at the possible BRST\
invariant counterterm $\Sigma _c$ which may affect the ultraviolet behavior
of the model. We recall that $\Sigma _c$ is an integrated local polynomial
with dimension bounded by two. Making use of the Ward identities established
in the previous section, $\Sigma _c$ is found to obey the conditions:

\begin{equation}
\frac{\delta \Sigma _c}{\delta b^c}={\cal G}^a\Sigma _c=0\;,  \label{stab-1}
\end{equation}

\begin{equation}
\left( \frac \delta {\delta \bar{c}^a}+\partial _\mu \frac \delta {\delta
\gamma _\mu ^a}\right) =0\,\,,  \label{stab-2}
\end{equation}

\begin{equation}
{\cal B}_\Sigma \Sigma _c=0\;,  \label{stab-3}
\end{equation}

\begin{equation}
{\cal F}_{\mu \nu }(x)\Sigma _c=0\;,  \label{stab-4}
\end{equation}
\begin{equation}
{\cal W}_\mu \Sigma _c=0\;.  \label{stab-5}
\end{equation}
From eq.(\ref{stab-1}) it follows that $\Sigma _c$ is independent from the
Lagrange multiplier $b^a$ and that it depends only from the differentiated
ghost $\partial _\mu c^a$ \cite{ps}. Moreover, due to the eq.(\ref{stab-2}),
the fields $\bar{c}^a$ and $\gamma _\mu ^a$ enter through the combination 
\cite{ps} 
\begin{equation}
\hat{\gamma}_\mu ^a\equiv \gamma _\mu ^a+\partial _\mu \bar{c}^a.
\end{equation}
Finally, from the conditions (\ref{stab-3}), (\ref{stab-4}) it turns out
that the most general BRST invariant counterterm can be written as

\begin{equation}
\Sigma _c=\Xi +{\cal B}_\Sigma \tilde{\Xi}\;,  \label{s-c}
\end{equation}
where 
\begin{equation}
\Xi =\eta \int d^2x\,\,\frac{\phi ^2}2,  \label{counter fi}
\end{equation}
with $\eta $ being an arbitrary parameter and $\tilde{\Xi}$ stands for an
integrated local polynomial with ghost number -1, representing the trivial
part of the cohomology of the operator ${\cal B}_\Sigma $. Observe that, due
to the eq.(\ref{stab-4}) and to the algebraic relations (\ref{c-a}), we have 
\begin{equation}
{\cal F}_{\mu \nu }(x){\cal B}_\Sigma \tilde{\Xi}={\cal B}_\Sigma {\cal F}%
_{\mu \nu }(x)\tilde{\Xi}=0\;,  \label{cond}
\end{equation}
from which it follows that

\begin{equation}
{\cal F}_{\mu \nu }(x)\tilde{\Xi}={\cal B}_\Sigma \Lambda _{\mu \nu }(x)\;,
\label{tr}
\end{equation}
for some local polynomial $\Lambda _{\mu \nu }(x)$ of ghost number -3. We
are left thus with a unique nontrivial BRST\ invariant counterterm given by
eq.(\ref{counter fi}). Expression (\ref{counter fi}) is physically
equivalent to the standard Yang-Mills counterterm $\int d^2x\,\,F^{a\mu \nu
}F_{\mu \nu }^a.$ This statement relies on the observation that the
auxiliary field $\phi ^a$ has in fact the role of $\varepsilon ^{\mu \nu
}F_{\mu \nu }^a$, as it is implied by the equations of motion. Also, it
should be observed that the coefficient $\eta $ in eq.(\ref{counter fi}) has
the meaning of a possible ultraviolet renormalization of the gauge coupling
constant $g$ compatible with the BRST invariance.

It remains now to impose the final constraint (\ref{stab-5}). Making use of
the algebraic relations (\ref{c-a}), it follows that

\begin{eqnarray}
{\cal W}_\mu {\cal B}_\Sigma \tilde{\Xi} &=&-{\cal B}_\Sigma {\cal W}_\mu 
\tilde{\Xi}+\left\{ {\cal B}_\Sigma ,{\cal W}_\mu \right\} \tilde{\Xi} 
\nonumber  \label{WBcounter} \\
&=&-{\cal B}_\Sigma {\cal W}_\mu \tilde{\Xi}+\int d^2x\xi ^\nu {\cal F}_{\nu
\mu }\tilde{\Xi}\;  \nonumber \\
&=&-{\cal B}_\Sigma \left( {\cal W}_\mu \tilde{\Xi}+\int d^2x\xi ^\nu
\Lambda _{\nu \mu }\right) \;,\;  \label{w-coun}
\end{eqnarray}
where use has been made of eq.(\ref{tr}) and of the fact that ${\cal B}%
_\Sigma \xi _\mu =0.\;$Therefore, from the requirement of invariance under
the Ward operator ${\cal W}_\mu $, we have

\begin{equation}
{\cal W}_\mu \Sigma _c={\cal W}_\mu \Xi -{\cal B}_\Sigma \left( {\cal W}_\mu 
\tilde{\Xi}+\int d^2x\xi ^\nu \Lambda _{\nu \mu }\right) =0\;,  \label{f-w-c}
\end{equation}
{\it i.e}.

\begin{equation}
-\eta \int d^2x\,\,\varepsilon _{\mu \nu }\hat{\gamma}^{a\nu }\phi ^a={\cal B%
}_\Sigma \left( {\cal W}_\mu \tilde{\Xi}+\int d^2x\xi ^\nu \Lambda _{\nu \mu
}\right) \;,  \label{eta-c}
\end{equation}
implying that $\int d^2x\,\,\varepsilon _{\mu \nu }\hat{\gamma}^{a\nu }\phi
^a$ is a trivial element of the integrated cohomology of ${\cal B}_\Sigma $.
However, it can be shown that this term cannot be actually cast in the form
of a pure ${\cal B}_\Sigma -$variation, as it identifies a nontrivial
element of the integrated cohomolgy of ${\cal B}_\Sigma $, namely

\begin{equation}
\int d^2x\,\,\varepsilon _{\mu \nu }\hat{\gamma}^{a\nu }\phi ^a\neq {\cal B}%
_\Sigma -{\rm variation\;.}  \label{nontriv}
\end{equation}
The only way out is thus 
\begin{equation}
\eta =0\;,  \label{zero-eta}
\end{equation}
meaning that there is no nontrivial BRST\ invariant counterterm compatible
with the vector Ward identity (\ref{w-i}), providing therefore a simple
algebraic understanding of the well known ultraviolet finiteness properties
of 2DYM\footnote{%
We shall not be concerned here with possible infrared pathologies of the
model, our aim being that of analyzing the ultraviolet region. A useful BRST
invariant infrared regularization could be introduced along the lines
developed in \cite{ir,bm} and already successfully applied in the case of
the 2D topological BF model \cite{bm}.}. It is useful remarking that the
algebraic proof of the absence of nontrivial counterterms given here follows
exactly the same lines of the proofs of the ultraviolet finiteness of the
topological models \cite{DGS e DLPS,GMS, algebraic,bm,wi}, emphasizing in
particular the pivotal role of the vector Ward identity (\ref{w-i}).

\vspace{5mm}

{\bf Acknowledgements}

The authors are grateful to J.A. Helay\"{e}l-Neto for helpful discussions
and very pertinent remarks. The Conselho Nacional de Desenvolvimento
Cient\'{\i}fico e Tecnol\'{o}gico CNPq-Brazil, the Funda\c {c}\~{a}o de
Amparo \`{a} Pesquisa do Estado do Rio de Janeiro (Faperj) and the SR2-UERJ
are acknowledged for the financial support.

\end{document}